\documentclass[sigconf,natbib=false,screen,noacm]{acmart}

\makeatletter
\def\@ACM@checkaffil{%
    \if@ACM@instpresent\else
    \fi
    \if@ACM@citypresent\else
    \fi
    \if@ACM@countrypresent\else
    \fi
}
\makeatother

\renewcommand\footnotetextcopyrightpermission[1]{} %

\usepackage[style=trad-abbrv, backend=biber, maxnames=1, minnames=1, doi=true,isbn=false,url=true , giveninits=true, sortcites]{biblatex}
\usepackage{graphicx}
\usepackage{booktabs}
\usepackage{hyperref}
\usepackage{cleveref}
\usepackage{siunitx}
\usepackage{amsthm}
\usepackage{multirow}
\usepackage{caption}
\usepackage{enumitem}
\usepackage{subcaption}
\usepackage{float}
\usepackage[defaultlines=3,all]{nowidow}

\newtheorem{example}{Example}

\graphicspath{{figures/}}

\begin{document}

\title[\phantom]{Hybrid Circuit Mapping:\\Leveraging the Full Spectrum of Computational Capabilities\\of Neutral Atom Quantum Computers}

\author[\phantom]{Ludwig Schmid$^\ast $\hspace{2.0em} Sunghye Park$^\dagger$ \hspace{2.0em} Seokhyeong Kang$^\dagger$ \hspace{2.0em} Robert Wille$^{\ast \, \ddagger}$ }
\affiliation{%
  \institution{$^\ast$Technical University of Munich, Germany}
  \city{}
  \country{}
}
\affiliation{%
  \institution{$^\dagger$Pohang University of Science and Technology, Korea}
  \city{}
  \country{}
}

\affiliation{%
  \institution{$^\ddagger$Software Competence Center Hagenberg GmbH, Hagenberg, Austria}
  \city{}
}
\email{{ludwig.s.schmid,robert.wille}@tum.de,{shpark96,shkang}@postech.ac.kr}

\begin{abstract}
Quantum computing based on \emph{Neutral Atoms}~(NAs) provides
a wide range of computational capabilities, encompassing 
  \mbox{high-fidelity} long-range interactions with native multi-qubit gates, and the ability to shuttle arrays of qubits.
   While previously these capabilities have been studied individually, we propose the first approach of a fast hybrid compiler to perform circuit mapping and routing based on both high-fidelity gate interactions and qubit shuttling.
   We delve into the intricacies of the compilation process when combining multiple capabilities and present effective solutions to address resulting challenges.
   The final compilation strategy is then showcased across various hardware settings, revealing its versatility, and highlighting potential fidelity enhancements achieved through the strategic utilization of combined gate- and shuttling-based routing.
   With the additional multi-qubit gate support for both routing capabilities, the proposed approach is able to take advantage of the full spectrum of computational capabilities offered by NAs.
\end{abstract}

\maketitle
\pagestyle{plain} 

\section{Introduction}\label{sec:introduction}
\emph{Neutral Atoms}~(NAs) have emerged as a compelling choice for universal quantum computing~\cite{saffmanQuantumInformationRydberg2010,saffmanQuantumComputingAtomic2016,henrietQuantumComputingNeutral2020,grahamMultiqubitEntanglementAlgorithms2022}, showcasing a broad spectrum of computational capabilities that encompass \mbox{high-fidelity}, long-range interactions between qubits with native multi-qubit gate support~\cite{grahamMultiqubitEntanglementAlgorithms2022,everedHighfidelityParallelEntangling2023,levineParallelImplementationHighFidelity2019}, and remarkable scalability~\cite{barredoAtombyatomAssemblerDefectfree2016,pauseSuperchargedTwodimensionalTweezer2023}. In response to these advantages, dedicated software solutions, such as compilers~\cite{bakerExploitingLongDistanceInteractions2021,liTimingAwareQubitMapping2023,patelGeyserCompilationFramework2022}, have been developed to optimize performance while adhering to hardware constraints.
Moreover, \mbox{\textcite{bluvsteinQuantumProcessorBased2022}} have demonstrated the ability to dynamically rearrange qubit arrays during computation with high fidelity.
Based on this experimental progress, further
compilation strategies~\cite{tanCompilingQuantumCircuits2023,nottinghamDecomposingRoutingQuantum2023,brandhoferOptimalMappingNearTerm2021} have explored the potential of using qubit shuttling for circuit mapping and routing, presenting it as a promising alternative to conventional approaches reliant on SWAP gate insertion.

Nevertheless, previous work individually only studied a single aspect or a subset of the full spectrum of the computational capabilities of NAs.
In particular, SWAP gate insertion and atom shuttling have been considered separately from each other.
While these separate studies of the mapping capabilities facilitate the initial understanding, they neglect potentially better solutions, arising from the combined use of both gate-based and shuttling-based mapping throughout the compilation process.

In this work, a 
hybrid compilation approach is proposed to explore the potential advantage of leveraging gate-based SWAP insertion \emph{and} shuttling-based atom rearrangements.
In particular, this entails the compilation task of mapping and routing a provided quantum circuit to NA hardware.
The resulting compilation process is able to choose between the two mapping capabilities for each gate individually within the circuit based on available hardware information.
Challenges arising from the simultaneous consideration of both mapping capabilities are discussed and 
corresponding solutions are proposed and integrated 
into a 
hybrid mapping process.
The heuristic approach employs two capability-specific cost functions, which are designed for rapid evaluation and consider additional information to enhance parallelism by incorporating commutation rules and look-ahead functionality.

The approach is evaluated on a set of benchmark circuits with up to $200$ qubits, considering different hardware information.
The evaluations demonstrate the ability of the proposed approach to correctly identify the preferred mapping capability of different hardware configurations.
In particular, it shows fidelity improvements for mixed hardware that is characterized by similar operation fidelities between entangling gates and shuttling.
Furthermore, the performed evaluations indicate a correlation between circuit structure and preferential mapping capability, offering new research questions, enabled by the hybrid mapping process.

Overall, the proposed hybrid approach is a further step to take advantage of the potential provided by NAs, offering new possibilities for mapping quantum circuits to hardware.
With the additional support for arbitrary-sized multi-qubit gates for both gate- and shuttling-based mapping, it represents the first proposal to leverage the full spectrum of computational capabilities offered by the NA hardware. The full code of the proposed approach, including evaluation data, is publicly available at~\cite{LsschmidMqtqmapHybridmapper} and will be integrated into the Munich Quantum Toolkit~(MQT)\footnote{\href{https://mqt.readthedocs.io/en/latest/}{https://mqt.readthedocs.io/en/latest/}} for general use.

The remainder of this paper is structured as follows:
\Cref{sec:background} provides a brief background on NAs and the task of circuit mapping employing gate- or shuttling-based capabilities.
In \Cref{sec:hybr-mapp-problem}, the hybrid compilation approach is introduced, by first discussing challenges and the corresponding solutions arising from the simultaneous use of both mapping capabilities, with the complete process overview and details discussed afterward.
In \Cref{sec:numer-eval}, we summarize the results of the numerical evaluations considering different hardware parameters and compiler settings, demonstrating the flexibility of the proposed approach and the potential for fidelity improvements.
Finally, \Cref{sec:summary-outlook} concludes the paper with a brief summary of the results of this work.

\section{Background}\label{sec:background}

\begin{figure*}[t]
     \centering
     \begin{subfigure}[t]{0.65\textwidth}
         \centering
  \includegraphics[width=\textwidth]{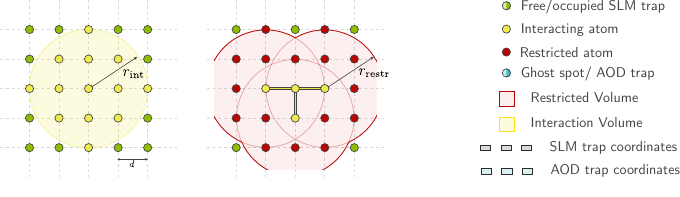}
         \caption{Interactions and restrictions for NA with \mbox{$r_\mathrm{int}=r_\mathrm{restr}=2d$} }
         \label{fig:multi_qubit_gates}
     \end{subfigure}
     \begin{subfigure}[t]{0.7\textwidth}
         \centering
         \includegraphics[width=\textwidth]{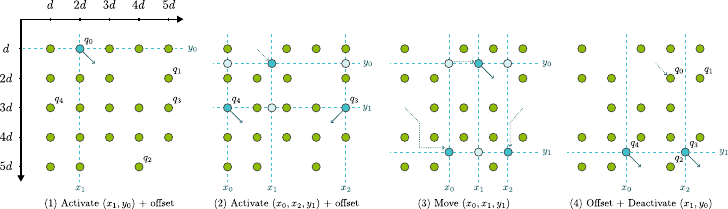}
         \caption{Qubit array shuttling with \emph{Acusto-Optic Deflectors} (AOD) parallelization constraints.}
         \label{fig:aod_parallelization}
     \end{subfigure}
  \caption{Computational capabilities of Neutral Atoms architectures }
\label{fig:comp-capability}
\end{figure*}

This section provides a brief summary of the computational capabilities of the NA platform~\cite{schmidComputationalCapabilitiesCompiler2023,henrietQuantumComputingNeutral2020,grahamMultiqubitEntanglementAlgorithms2022,saffmanQuantumInformationRydberg2010,saffmanQuantumComputingAtomic2016} followed by a review of the compilation task of mapping quantum circuits onto NA hardware.

\subsection{Neutral Atom Computational Capabilties}
\label{sec:neutr-atom-comput-capab}
In NAs, to realize a computational register, atoms are stored in optical dipole traps, such as optical lattices or optical tweezer traps.
These are created by interfering laser beams that create an array of potential valleys, effectively trapping the atoms at specific coordinates.
Within this work, we assume these traps to lay on a regular square lattice with lattice constant $d$.
Common atom species include alkali or alkaline-earth-like atoms such as Rb or Sr.
The computational states can be encoded in long-lived atomic states such as hyperfine- or nuclear spin-states, after laser-cooling the atoms down to their motional ground states.
Single qubit gates are then realized by laser pulses on individually addressed qubits or the whole register using globally applied laser beams.
Multi-qubit gates are realized using the Rydberg blockade effect to introduce a phase shift conditioned on the qubit states of the nearby excited atoms~\mbox{\cite{levineParallelImplementationHighFidelity2019,everedHighfidelityParallelEntangling2023}}.

This theoretically allows realizations of $m$-qubit \mbox{high-fidelity} multi-controlled $\mathrm{C}_{m-1} \mathrm{Z}$ phase gates.
For these gates to be executable, all participating qubits need to be within a certain \emph{\mbox{interaction} radius} $r_{\mathrm{int}}$ to each other, where $r_{\mathrm{int}}$ depends on the atom species and the chosen Rydberg state.
To reduce crosstalk between gates, parallel execution is only possible if qubits corresponding to different gates keep a distance of at least the \emph{restriction radius} $r_{\mathrm{restr}} \geq r_{\mathrm{int}}$ to all qubits that execute another multi-qubit gate simultaneously. The resulting region is referred to as \emph{restricted volume}.
\begin{example}
  The yellow region in \Cref{fig:multi_qubit_gates} indicates the possible interaction candidates for multi-qubit gates for $r_{\mathrm{int}}=r_{\mathrm{restr}}=2 d$.
  The red region corresponds to the restricted volume for other multi-qubit gates to be executed simultaneously.
\end{example}

In addition to these long-range interactions (for large $r_\mathrm{int}$), NAs provide the capability to \emph{shuttle} arrays of trapped atoms~\cite{bluvsteinQuantumProcessorBased2022}.
To this end, the qubits are loaded from the static \emph{\mbox{Spatial} Light Modulator}~(SLM) traps into a 2D \emph{Acousto-Optic Deflector}~(AOD).
The AOD can be described by the corresponding x (column) and y (row) coordinates of the deflected laser, where each intersection defines a potential trap.
Using more than one row and column allows for the shuttling of multiple qubits at once, each trapped in one of the intersecting coordinates.
Each row/column coordinate can be activated, moved, and finally deactivated.
This capability allows for arbitrary rearrangements of the atoms according to the two following constraints:

First, columns and rows are not allowed to cross each other, i.e. the ordering of the rows/columns always remains the same.
To move beyond another trapped qubit the first has to be released back to a static SLM trap by deactivating the corresponding row and/or column and, then, cross as a second step only.
Secondly, empty AOD intersections still represent a potential trap, disturbing qubits unintentionally when hovering or passing over them.
The effect of these \emph{ghost spots} can be circumvented by loading the qubits sequentially into the AOD traps, each with an additional offset movement to prevent the ghost spots from hovering over other qubits at any time.

\begin{example}
\label{ex:aod}
  \Cref{fig:aod_parallelization} illustrates a scenario where $q_{0}$ needs to be shuttled to $q_{1}$ and $q_{3},q_{4}$ to $q_{2}$.
  As a first step (1), $q_{0}$ is loaded by activating AOD coordinates at $x_{1} = 2d$ and $y_{0}=d$.
  To prevent problematic ghost spots with the following activations, a small offset move is applied to the coordinates.
  Then (2), $q_{3}$ and $q_{4}$ can be loaded simultaneously in the same row by activating $x_{0} = d$, $x_{2} = 5d$ and row $y_{1} = 3d$.
  Note that due to the previous offset, all resulting ghost spots (light blue) are in the empty inter-qubit regions.
  The qubits can then be moved (3) to their destinations without crossing any other AOD coordinate and placed sequentially using again offset movements (4).
\end{example}

\subsection{Quantum Circuit Mapping}
\label{sec:quant-circ-mapp}
The task of \emph{circuit mapping} consists of assigning circuit qubits \mbox{$\mathbf{Q} = \{q_{i}\}_{i = 0, \dots, n-1}$} to a set of physical qubits \mbox{$\mathbf{P} = \{Q_{a}\}_{a=0, \dots, N-1}$}.
In addition, for NAs the mapping task becomes two-folded as the set of \emph{physical qubits} $\mathbf{P}$ also has to be assigned to the possible \emph{trap coordinates} \mbox{$\mathbf{C} = \{C_{\alpha}\}_{\alpha=0, \dots, \mu}$}.
We assume \mbox{$\mu = l^{2} -1 > m \geq n$} with a non-zero number of unoccupied coordinates on a regular $l \times l$ lattice.
As a result, one has to consider two mapping steps: first, the \emph{qubit mapping} $f_\mathrm{q}$ of assigning circuit qubits to physical hardware qubits, represented by trapped atoms.
Secondly, the \emph{atom mapping} $f_\mathrm{a}$ of assigning the physical qubits to the coordinates.

Given both mappings, one can define the \emph{connectivity graph} $G = (\mathbf{P}, \mathbf{E})$, where $\mathbf{E}$ contains all pairs of physical qubits that can interact with each other, i.e.,~$(Q_{a},Q_{b}) \in \mathbf{E} \Leftrightarrow d(Q_{a},Q_{b}) \leq r_{\mathrm{int}}$ with $d$ the Euclidean distance. The set of qubits that are connected to $Q$ are referred to as \emph{vicinity} $V_{r_\mathrm{int}}(Q)$ and $K_{r_\mathrm{int}}=|V_{r_\mathrm{int}}(Q)|$ as its \emph{coordination number}.

\begin{figure}[t]
  \centering
  \includegraphics[width=1\columnwidth]{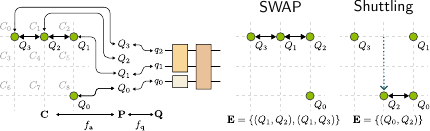}
  \caption{ \textbf{Qubit and atom mapping}}
\label{fig:double-mapping}
\end{figure}

\begin{example}
  The illustration in \Cref{fig:double-mapping} shows a $3\times3$ lattice with $|\mathbf{C}|=9$ SLM traps as well as $|\mathbf{P}|=4$ physical and $|\mathbf{Q}|=3$ circuit qubits to be mapped.
  The arrows indicate both qubit \mbox{mapping $f_\mathrm{q}$} as well as an atom mapping \mbox{$f_\mathrm{a}$}.
  For $r_\mathrm{int}=d$ this results in $\mathbf{E} = \{(Q_1,Q_2), (Q_2,Q_3)\}$ and, e.g., $V(Q_2) = \{Q_1,Q_3\}$.
\end{example}

Due to the limited connectivity between all qubits, in the following step, $G$ needs to be updated throughout circuit execution such that whenever a gate is executed, the corresponding physical qubits are in $\mathbf{E}$.
This is referred to as \emph{circuit routing} and is typically done by inserting SWAP gates into the circuit, swapping the qubit mapping assignment of two hardware qubits, and, this way, modifying $G$.
On hardware, the SWAP gates can be realized using 3 CX gates or equivalently 3 CZ gates with additional single qubit rotations.
In the last decade, multiple software tools to solve this problem for superconducting hardware have been developed, trying to minimize the number of SWAP gates~\cite{liTacklingQubitMapping2019,zulehnerEfficientMethodologyMapping2019,tanOptimalLayoutSynthesis2020,cowtanQubitRoutingProblem2019}.
For NAs, there are solutions taking into account variable interaction and restriction radii~\cite{bakerExploitingLongDistanceInteractions2021} as well as time-aware routing~\cite{liTimingAwareQubitMapping2023}.

On the other hand, the shuttling capability of NAs additionally allows for changing the atom mapping by physically moving atoms to another lattice coordinate and modifying $G$ this way.
This has the advantage of not introducing additional error-prone CZ gates, but, depending on the hardware setup, may be significantly slower than SWAP gate insertion.
Recent work has studied the potential of this novel mapping strategy, considering optimal solutions for a \mbox{shuttling-only} hardware setup by \textcite{tanCompilingQuantumCircuits2023} and another shuttling-only heuristic routing algorithm by \textcite{nottinghamDecomposingRoutingQuantum2023}.
The latter allows the crossing of AOD rows/columns which allows the reduction of the shuttling problem to be tackled in a similar way to the SWAP gate insertion and can, therefore, not be compared directly to the shuttling constraints considered in this work.

\begin{example}
  Considering again \Cref{fig:double-mapping}, applying $\mathrm{SWAP}(Q_{1},Q_{2})$, substitutes $(Q_{2},Q_{3})$ with $(Q_{1},Q_{3})$ in $\mathbf{E}$. Using shuttling, one can also modify $G$ by placing $Q_{2}$ at $C_{7}$ and, this way resulting in $\mathbf{E} = \{(Q_0,Q_2)\}$.
\end{example}

To evaluate the mapping results across different capabilities, common figures of merit such as SWAP gate count are no longer valid measures.
A possible alternative~\cite{schmidComputationalCapabilitiesCompiler2023} is the \emph{approximate success probability} $P$ defined as
\begin{equation}
  \label{eq:success-prob}
  P = \exp \left( - \dfrac{t_{\mathrm{idle}}}{T_{\mathrm{eff}}} \right) \prod_{O} \mathcal{F}_{O} \, , \quad T_\mathrm{eff} = \frac{T_1 T_2}{T_1 + T_2} \, ,
\end{equation}
where the product runs over all circuit operations, $T_1$, $T_2$ are the coherence times of the system, and $\mathcal{F}_{O} \in [0,1] $ is a measure for the average operation fidelity of operation $O$.
The \emph{total idle time} $t_{\mathrm{idle}}$ is defined as $t_{\mathrm{idle}} = n\cdot T - \sum_{O} t_{O}$ with $t_{O}$ the execution time of operation $O$ and $T$ is the total circuit runtime after scheduling, according to the constraints reviewed in \Cref{sec:neutr-atom-comput-capab}.

\section{Hybrid Mapping Problem}
\label{sec:hybr-mapp-problem}
Considering the full spectrum of capabilities of NAs, as reviewed above, results in a two-fold mapping problem.
First, in the \emph{gate-based mapping} step, one can route gates that are not connected trivially by modifying the qubit mapping using SWAP gate insertion.
Secondly, using \emph{shuttling-based mapping}, the qubit mapping remains unaltered but the connectivity graph is modified by moving atoms to a new trap coordinate and, therefore, acting on the atom mapping.
Recent approaches~\cite{bakerExploitingLongDistanceInteractions2021, tanCompilingQuantumCircuits2023, nottinghamDecomposingRoutingQuantum2023, liTimingAwareQubitMapping2023} only considered one of the two cases separately, leaving untouched potential room for improvement.
Motivated by that, this work proposes a compilation process that utilizes the two capabilities in a hybrid fashion, trying to employ the most suitable method to achieve the required connectivity.
This imposes multiple challenges as illustrated in the following \Cref{sec:challanges}, with the corresponding solutions brought together to the proposed hybrid compilation approach reviewed in \Cref{sec:comp-proc-overview}. Finally, \Cref{sec:impl-details} provides technical details such as the employed cost functions.

\subsection{Challenges and Proposed Solutions}
\label{sec:challanges}

\subsubsection{Increased Search Space}
\label{sec:increased-search}
The potential utilization of both mapping capabilities significantly increases the number of possible operations during the mapping process.
This amplified search space may allow finding better solutions but it also imposes a challenge on the compiler that needs to decide between all possibilities.
For the gate-based mapping, the set of possible SWAP operations corresponds to the union of possible SWAPs within $r_{\mathrm{int}}$ for all current gate qubits.
In the worst case, this results in $\mathcal{O}(n K_{r_{\mathrm{int}}} / 2)$, where $n$ is the number of circuit qubits and $K_{r_{\mathrm{int}}}$ is the coordination number.
Since, as for many circuits, only a subset of all qubits participate in the next executable gates, this number will be significantly lower in most practical cases.

For shuttling-based mapping on the other hand, potentially any of the $m$ physical atoms can be moved to any of the $\mu - m$ unoccupied coordinates.
If one, furthermore, considers the possibility of moving away atoms from certain coordinates to free the space, effectively any possible rearrangement of qubits is possible.
This leads to $\mathcal{O}(N |C|)$ potential shuttling operations at each step, making a full search space exploration unfeasible.

To reduce the search space, we only consider qubit movements that move the qubit directly in the vicinity of one of the other gate qubits.
This can be done directly if there is an unoccupied coordinate.
Otherwise, we select one of the nearby qubits to be moved away, resulting in a chain of two consecutive movements.

\begin{example}
\label{ex:search-space}
  \Cref{fig:challenge-inc-seachr-space} illustrates the considered search space to connect two qubits for $r_{\mathrm{int}}=\sqrt{2}d$.
  Shown are the possible operations for the three cases: (1) gate-based SWAPS, (2) directly shuttling to one of the free coordinates, and (3) requiring a move-away operation beforehand.
\end{example}

\subsubsection{Mapping conflicts}
\label{sec:double-mapp-probl}
\begin{figure}[t]
     \centering
     \begin{subfigure}[t]{0.99\columnwidth}
  \centering
  \includegraphics[width=\columnwidth]{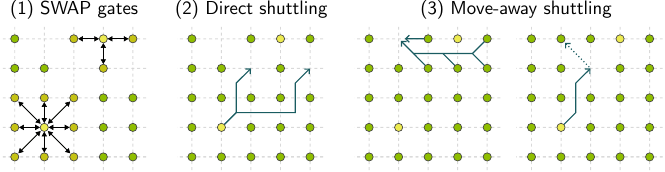}
  \caption{Increased search space: Considered operations.}
\label{fig:challenge-inc-seachr-space}
\end{subfigure}

     \begin{subfigure}[t]{0.9\columnwidth}
         \centering
  \includegraphics[width=\columnwidth]{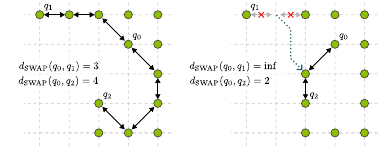}
  \caption{Mapping conflicts: Distance change by shuttling.}
\label{fig:challenge-double-mapping}
     \end{subfigure}
     \hfill
     \begin{subfigure}[t]{0.7\columnwidth}
         \centering
  \includegraphics[width=\columnwidth]{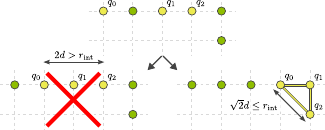}
  \caption{Multi-qubit gate-based mapping: Positioning.}
\label{fig:challenge-multi-qubit}
     \end{subfigure}
     \caption{Challenges and proposed solutions.}
     \vspace{-4mm}
\end{figure}

As both, gate- and shuttling-based mapping directly affect the connectivity graph $G$, they can unintentionally conflict with each other.
In particular, shuttling qubits away does not only affect the mapping of the moved qubits and their vicinity but it may also influence the optimal SWAP path for other qubits.
\begin{example}
  \Cref{fig:challenge-double-mapping} illustrates the case, where a shuttling operation affects the number of necessary SWAPs $d_\mathrm{SWAP}$ also for gates that do not directly act on the moved qubit.
  In this case, the distance between $q_{0}$ and $q_{2}$ is reduced by the shuttling operation, while the connection $q_{0}$ to $q_{1}$ becomes impossible as the previous connection no longer exists.
\end{example}

\subsubsection{Multi-Qubit Gate-based Mapping}
\label{sec:multi-qubit-gate}
As discussed in \Cref{sec:neutr-atom-comput-capab}, we assume multi-qubit gates ($m \geq 3$) to be executable if all gate qubits are within the interaction radius of each other.
This allows for different geometric arrangements of the qubits such as bulky clusters or other more spacious geometries for large $r_{\mathrm{int}}$.
While for static architectures the geometry is not a problem, the dynamic rearrangement of the qubits may result in a situation where there are not enough qubits or with the wrong geometric arrangement.
As a result, one can not employ a distance-based cost function that directly drives qubits close to each other but it actually has to be checked if the required geometric realization is possible and, if yes, where.
This can be done with a breadth-first search, starting simultaneously from all gate qubits.
If it is not possible to find a corresponding set of hardware qubits to execute the multi-qubit gate with SWAP gates, shuttling-based mapping has to be employed instead.
\begin{example}
  Assuming $r_{\mathrm{int}}=\sqrt{2}d$, \Cref{fig:challenge-multi-qubit} illustrates a case where a distance-only ``move-together'' approach for multi-qubit gates mapping will fail.
  Due to the small $r_{\mathrm{int}}$, the execution of the gate between qubits $q_{0}$,$q_{1}$ and $q_{2}$ requires a rectangular arrangement of the qubits.
  Therefore, driving the qubits close to each other results in a dead end.
  Instead, $G$ has to be parsed to find the position with suitable geometry, as shown on the right.
\end{example}

\subsection{Resulting Overall Mapping Process}

\begin{figure}[t]
\vspace{-4mm}
  \centering
  \includegraphics[width=0.99\columnwidth]{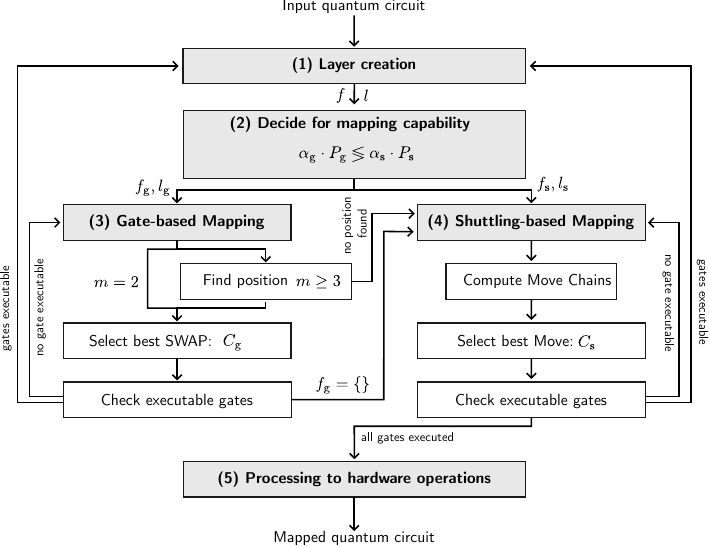}
  \caption{ \textbf{Resulting hybrid mapping process}}
\label{fig:hybrid-compil-process}
\vspace{-4mm}
\end{figure}

Taking into account the discussed challenges and the respective proposed solution ideas, we propose the following overall hybrid mapping process to leverage gate-based and shuttling-based mapping.
The process can be described as five major building blocks, illustrated graphically in \Cref{fig:hybrid-compil-process}.
\begin{enumerate}[wide,labelindent=0pt, nosep]
  \item \textbf{Layer creation:} Creates a frontier layer $f$ of gates that can be executed next, taking into account commutation rules. An additional lookahead layer $l$ contains gates following the frontier layer up to a certain lookahead depth.
  \item \textbf{Decide for mapping capability:} For each gate, an estimate of the required number of SWAPs and shuttling operations is computed.
        Based on this estimate, an \emph{approximate success probability}~$P_\mathrm{g}$ and $ P_\mathrm{s}$ according to \Cref{eq:success-prob} is derived in both cases.
        After weighing the outcomes with $\alpha_{\mathrm{g}}$ and $\alpha_{\mathrm{s}}$ respectively the gate is assigned to the respective front/lookahead layers of gate-based $f_{\mathrm{g}}$, $l_{\mathrm{g}}$ or shuttling-based $f_{\mathrm{s}}, l_{\mathrm{s}}$ mapping.
  \item \textbf{Gate-based Mapping:} Computes the best SWAP for all gates in $f_{\mathrm{g}}$ based on a distance-dependent cost function discussed in \Cref{sec:gate-based-routing}.
  For $m \geq 3$, a suitable position on the graph has to be found (use shuttling otherwise).
  This is repeated until at least one gate can be executed, resulting in updating the layers.
  \item \textbf{Shuttling-based Mapping:} Computes chains of possible shuttling operations and chooses the best one according to the cost function discussed in \Cref{sec:shuttl-based-rout}.
  This is again repeated until one of the gates can be executed in which case all layers are again updated.
  To prevent interference with the gate-based mapping, $f_{\mathrm{s}}$ is only considered if $f_{\mathrm{g}}$ is empty, meaning all necessary SWAP gates have been already applied.
  \item \textbf{Processing to hardware operations:} Finally, the SWAP gates are decomposed to natively supported CZ and single-qubit gates.
  Simultaneously, the shuttling operations are scheduled in parallel according to the AOD constraints and converted to native AOD operations, entailing AOD activation, deactivation, and movements of the AOD coordinates (see \Cref{ex:aod}).
\end{enumerate}
The resulting output can then be scheduled to compute characteristics such as the total circuit execution time $T$, or the total qubit idle time $t_\mathrm{idle}$.
This scheduling step also takes into account the restriction constraint regarding the parallel execution of \mbox{multi-qubit} gates discussed in \Cref{sec:neutr-atom-comput-capab}.

In the following, some parts of the compilation process are discussed in more detail, in particular, the employed cost functions and multi-qubit mapping for $m \geq 3$. 
The full code, including evaluation data is publicly available as \mbox{open-source} software at~\cite{LsschmidMqtqmapHybridmapper}.

\label{sec:comp-proc-overview}

\subsection{Implementation Details}
\label{sec:impl-details}

\subsubsection{Gate-based Routing}
\label{sec:gate-based-routing}
According to the previous discussions in \Cref{sec:multi-qubit-gate}, for the gate-based mapping it is necessary to differentiate between two and multi-qubit gates with $m \geq 3$.
The former can be swapped close to each other, while the latter requires the search for a suitable geometric position to execute the gate.
This position is found by employing a breadth-first search on $G$, starting simultaneously from all gate qubits and choosing the resulting position that requires the least number of SWAPS.
Assuming a position~$P_g$ is found for gate $g$, the following cost function based on the two-qubit cost of \textcite{liTacklingQubitMapping2019} is evaluated for each SWAP candidate $S$:
\begin{align}
  \label{eq:cost-gate-based}
C_{\mathrm{g}}(S) &= e^{- \lambda^t t(S) } \left[ C^{f}_{\mathrm{g}}(S) + w_{l} \cdot C^{l}_{\mathrm{g}}(S) \right] \\
  C^{f}_{\mathrm{g}}(S) &= \sum_{g \in f_{\mathrm{g}}\,,\, m = 2} \Delta d_{\mathrm{SWAP}}(S,g) + \sum_{g \in f_{\mathrm{g}}\, , \, m \geq 2} \Delta d_{\mathrm{SWAP}}(S,P_g) \, ,
\end{align}
where $\Delta d_{\mathrm{SWAP}}(S,g|P_g)$ returns the difference in the number of SWAPs for \mbox{gate $g$} or position $P_g$ after the application of $S$.
$C_{\mathrm{g}}$ computes the cost for the front and lookahead layer, with a \emph{lookahead weighting factor}~$w_{l}$.
This sum is then weighted with an exponential factor constituted by a decay rate $\lambda^t$ and a ``last used'' integer $t(S)$ to favor the use of SWAPs acting on different qubits to increase parallelism.
In difference to \textcite{liTacklingQubitMapping2019}, $t(S)$ also takes into account restricted qubits due to the NA-specific constraint of $r_{\mathrm{restr}}$.
By choosing different values for $\lambda \in \mathbb{R}^{+}$, this allows a continuous control between high parallelism and minimizing the absolute number of SWAP gates.
Depending on the given hardware setup, one of the two figures of merits may be favorable over the other~\cite{schmidComputationalCapabilitiesCompiler2023}, and, tuning $\lambda$ allows for more hardware adaptive mapping compared to other NA compilers such as \textcite{bakerExploitingLongDistanceInteractions2021} or \textcite{liTimingAwareQubitMapping2023}.

\begin{table*}[h]
\caption{ \textbf{Mapping Results for different NA hardware settings with different compilation strategies}}
\begin{subtable}{0.75\textwidth}
\centering
\caption{\footnotesize Mapping results}
\label{tab:table1_a}
\resizebox{1.0\textwidth}{!}{
\begin{tabular}{ccc|rrrrrrrrrrrrc|}
\cline{4-16}
 &  & \multicolumn{1}{r|}{} & \multicolumn{13}{c|}{\textbf{Compiler Settings}} \\ %
 &  &  & \multicolumn{4}{c}{(A) Shuttling-based only} & \multicolumn{4}{c}{(B) Gate-based only} & \multicolumn{5}{c|}{(C) Proposed Hybrid Approach} \\ %
 &  &  & \multicolumn{1}{c}{$\Delta$CZ} & \multicolumn{1}{c}{$\Delta T$ [\SI{}{\micro\s}]} & \multicolumn{1}{c}{$\delta \mathcal{F}$} & \multicolumn{1}{c}{RT {[\SI{}{\s}]}} & \multicolumn{1}{c}{$\Delta$CZ} & \multicolumn{1}{c}{$\Delta T$ [\SI{}{\micro\s}]} & \multicolumn{1}{c}{$\delta \mathcal{F}$} & \multicolumn{1}{c}{RT {[\SI{}{\s}]}} & \multicolumn{1}{c}{$\Delta$CZ} & \multicolumn{1}{c}{$\Delta T$ [\SI{}{\micro\s}]} & \multicolumn{1}{c}{$\delta \mathcal{F}$} & \multicolumn{1}{c}{RT [\SI{}{\s}]} & $\alpha$ \\ \hline
\multicolumn{1}{|r}{\multirow{18}{*}{\rotatebox[origin=c]{90}{\textbf{Hardware Settings}}}} & \multicolumn{1}{c|}{\multirow{6}{*}{\rotatebox[origin=c]{90}{(1) Shuttling}}} & \textit{graph} & \multicolumn{1}{r}{\textbf{0}} & \multicolumn{1}{r}{6.3} & \multicolumn{1}{r}{\textbf{0.45}} & \multicolumn{1}{r|}{4.9} & \multicolumn{1}{r}{1086} & \multicolumn{1}{r}{\textbf{0.1}} & \multicolumn{1}{r}{7.37} & \multicolumn{1}{r|}{4.9} & \multicolumn{1}{r}{\textbf{0}} & \multicolumn{1}{r}{6.3} & \multicolumn{1}{r}{\textbf{0.45}} & \multicolumn{1}{r|}{4.9} & \multirow{12}{*}{1} \\
\multicolumn{1}{|r}{} & \multicolumn{1}{c|}{} & \textit{qft} & \multicolumn{1}{r}{\textbf{0}} & \multicolumn{1}{r}{357.0} & \multicolumn{1}{r}{\textbf{25.20}} & \multicolumn{1}{r|}{60.6} & \multicolumn{1}{r}{12141} & \multicolumn{1}{r}{\textbf{5.4}} & \multicolumn{1}{r}{88.47} & \multicolumn{1}{r|}{60.7} & \multicolumn{1}{r}{\textbf{0}} & \multicolumn{1}{r}{357.0} & \multicolumn{1}{r}{\textbf{25.20}} & \multicolumn{1}{r|}{60.4} &  \\
\multicolumn{1}{|r}{} & \multicolumn{1}{c|}{} & \textit{qpe} & \multicolumn{1}{r}{\textbf{0}} & \multicolumn{1}{r}{313.5} & \multicolumn{1}{r}{\textbf{22.79}} & \multicolumn{1}{r|}{61.7} & \multicolumn{1}{r}{11928} & \multicolumn{1}{r}{\textbf{3.5}} & \multicolumn{1}{r}{87.23} & \multicolumn{1}{r|}{61.4} & \multicolumn{1}{r}{\textbf{0}} & \multicolumn{1}{r}{313.5} & \multicolumn{1}{r}{\textbf{22.79}} & \multicolumn{1}{r|}{61.7} &  \\
\multicolumn{1}{|r}{} & \multicolumn{1}{c|}{} & \textit{bn} & \multicolumn{1}{r}{\textbf{0}} & \multicolumn{1}{r}{11.0} & \multicolumn{1}{r}{\textbf{1.21}} & \multicolumn{1}{r|}{0.3} & \multicolumn{1}{r}{522} & \multicolumn{1}{r}{\textbf{0.1}} & \multicolumn{1}{r}{4.06} & \multicolumn{1}{r|}{0.3} & \multicolumn{1}{r}{\textbf{0}} & \multicolumn{1}{r}{11.0} & \multicolumn{1}{r}{\textbf{1.21}} & \multicolumn{1}{r|}{0.3} &  \\
\multicolumn{1}{|r}{} & \multicolumn{1}{c|}{} & \textit{call} & \multicolumn{1}{r}{\textbf{0}} & \multicolumn{1}{r}{16.6} & \multicolumn{1}{r}{\textbf{1.27}} & \multicolumn{1}{r|}{35.8} & \multicolumn{1}{r}{1146} & \multicolumn{1}{r}{\textbf{0.3}} & \multicolumn{1}{r}{8.10} & \multicolumn{1}{r|}{35.5} & \multicolumn{1}{r}{\textbf{0}} & \multicolumn{1}{r}{16.6} & \multicolumn{1}{r}{\textbf{1.27}} & \multicolumn{1}{r|}{35.5} &  \\
\multicolumn{1}{|r}{} & \multicolumn{1}{c|}{} & \textit{gray} & \multicolumn{1}{r}{\textbf{0}} & \multicolumn{1}{r}{5.8} & \multicolumn{1}{r}{\textbf{0.44}} & \multicolumn{1}{r|}{0.5} & \multicolumn{1}{r}{483} & \multicolumn{1}{r}{\textbf{0.1}} & \multicolumn{1}{r}{3.34} & \multicolumn{1}{r|}{0.5} & \multicolumn{1}{r}{\textbf{0}} & \multicolumn{1}{r}{5.8} & \multicolumn{1}{r}{\textbf{0.44}} & \multicolumn{1}{r|}{0.5} &  \\ \cline{3-15}
\multicolumn{1}{|r}{} & \multicolumn{1}{c|}{\multirow{6}{*}{\rotatebox[origin=c]{90}{(2) Gate}}} & \textit{graph} & \textbf{0} & 11.8 & 0.94 & \multicolumn{1}{r|}{2.8} & 324 & \textbf{0.0} & \textbf{0.10} & \multicolumn{1}{r|}{0.4} & 324 & \textbf{0.0} & \textbf{0.10} & \multicolumn{1}{r|}{0.4} &  \\
\multicolumn{1}{|r}{} & \multicolumn{1}{c|}{} & \textit{qft} & \textbf{0} & 465.7 & 32.15 & \multicolumn{1}{r|}{70.2} & 2733 & \textbf{2.3} & \textbf{0.96} & \multicolumn{1}{r|}{17.4} & 2733 & \textbf{2.3} & \textbf{0.96} & \multicolumn{1}{r|}{17.4} &  \\
\multicolumn{1}{|r}{} & \multicolumn{1}{c|}{} & \textit{qpe} & \textbf{0} & 563.4 & 39.80 & \multicolumn{1}{r|}{50.2} & 3021 & \textbf{2.3} & \textbf{1.05} & \multicolumn{1}{r|}{17.4} & 3021 & \textbf{2.3} & \textbf{1.05} & \multicolumn{1}{r|}{17.5} &  \\
\multicolumn{1}{|r}{} & \multicolumn{1}{c|}{} & \textit{bn} & \textbf{0} & 29.9 & 2.12 & \multicolumn{1}{r|}{0.6} & 132 & \textbf{0.0} & \textbf{0.04} & \multicolumn{1}{r|}{0.1} & 132 & \textbf{0.0} & \textbf{0.04} & \multicolumn{1}{r|}{0.1} &  \\
\multicolumn{1}{|r}{} & \multicolumn{1}{c|}{} & \textit{call} & \textbf{0} & 28.9 & 2.07 & \multicolumn{1}{r|}{28.2} & 309 & \textbf{0.1} & \textbf{0.10} & \multicolumn{1}{r|}{26.4} & 309 & \textbf{0.1} & \textbf{0.10} & \multicolumn{1}{r|}{26.4} &  \\
\multicolumn{1}{|r}{} & \multicolumn{1}{c|}{} & \textit{gray} & \textbf{0} & 9.2 & 0.64 & \multicolumn{1}{r|}{0.2} & 72 & \textbf{0.0} & \textbf{0.02} & \multicolumn{1}{r|}{0.0} & 72 & \textbf{0.0} & \textbf{0.02} & \multicolumn{1}{r|}{0.0} &  \\ \cline{3-16} 
\multicolumn{1}{|r}{} & \multicolumn{1}{c|}{\multirow{6}{*}{\rotatebox[origin=c]{90}{(3) Mixed}}} & \textit{graph} & \textbf{0} & 9.9 & \textbf{0.59} & \multicolumn{1}{r|}{4.6} & 846 & \textbf{0.1} & 2.54 & \multicolumn{1}{r|}{0.4} & \textbf{0} & 9.9 & \textbf{0.59} & \multicolumn{1}{r|}{4.6} & 0.95 \\
\multicolumn{1}{|r}{} & \multicolumn{1}{c|}{} & \textit{qft} & \textbf{0} & 607.3 & 36.08 & \multicolumn{1}{r|}{67.9} & 8898 & \textbf{4.4} & 27.24 & \multicolumn{1}{r|}{18.3} & 6867 & 54.1 & \textbf{24.09} & \multicolumn{1}{r|}{24.0} & 1.04 \\
\multicolumn{1}{|r}{} & \multicolumn{1}{c|}{} & \textit{qpe} & \textbf{0} & 613.7 & 36.51 & \multicolumn{1}{r|}{59.6} & 7980 & \textbf{2.9} & 24.41 & \multicolumn{1}{r|}{20.1} & 6987 & 26.9 & \textbf{22.82} & \multicolumn{1}{r|}{22.9} & 1.06 \\
\multicolumn{1}{|r}{} & \multicolumn{1}{c|}{} & \textit{bn} & \textbf{0} & 21.6 & 1.28 & \multicolumn{1}{r|}{0.3} & 492 & \textbf{0.2} & 1.47 & \multicolumn{1}{r|}{0.1} & 78 & 10.0 & \textbf{0.83} & \multicolumn{1}{r|}{0.2} & 1.01 \\
\multicolumn{1}{|r}{} & \multicolumn{1}{c|}{} & \textit{call} & \textbf{0} & 34.3 & 2.04 & \multicolumn{1}{r|}{37.9} & 693 & \textbf{0.2} & 2.09 & \multicolumn{1}{r|}{35.0} & 363 & \textbf{11.4} & \textbf{1.77} & \multicolumn{1}{r|}{36.3} & 1.01 \\
\multicolumn{1}{|r}{} & \multicolumn{1}{c|}{} & \textit{gray} & \textbf{0} & 8.6 & 0.52 & \multicolumn{1}{r|}{0.5} & 285 & \textbf{0.1} & 0.86 & \multicolumn{1}{r|}{0.1} & 66 & 4.5 & \textbf{0.46} & \multicolumn{1}{r|}{0.2} & 0.99 \\ \hline
\end{tabular}
}
\end{subtable}
\begin{subtable}{0.24\textwidth}
\caption{\footnotesize Benchmark descriptions}
\label{tab:table1_b}
\centering
\begin{subtable}{1.0\textwidth}
\resizebox{1.0\textwidth}{!}{
\begin{tabular}{|c|cccc|}
\hline
Name & \multicolumn{1}{c}{$n$} & \multicolumn{1}{c}{nCZ} & \multicolumn{1}{c}{n$\mathrm{C}_2 \mathrm{Z}$} & \multicolumn{1}{c|}{n$\mathrm{C}_3 \mathrm{Z}$} \\ \hline
\textit{graph} & 200 & 215 & 0 & 0\\
\textit{qft} & 200 & 9998 & 0 & 0\\
\textit{qpe} & 200 & 10340 & 0 & 0\\
\textit{bn} & 48 & 133 & 87 & 0\\
\textit{call} & 25 & 0 & 192 & 56\\
\textit{gray} & 33 & 0 & 62 & 0\\ \hline
\end{tabular}
}
\vspace{-0.065cm}
\end{subtable}
\begin{subtable}{1.0\textwidth}
\caption{\footnotesize Hardware settings}
\label{tab:table1_c}
\resizebox{1.0\textwidth}{!}{
\begin{tabular}{|c|ccc|}
\hline
Paramters & \multicolumn{1}{c|}{Shuttling} & \multicolumn{1}{c|}{Gate} & Mixed \\ \hline
$r_{\mathrm{int}} = r_{\mathrm{rest}}$ & \multicolumn{1}{c|}{2} & \multicolumn{1}{c|}{4.5} & 2.5 \\ \hline
$\mathcal{F}_{\mathrm{CZ}}$ & \multicolumn{1}{c|}{0.994} & \multicolumn{1}{c|}{0.9995} & 0.995 \\
$\mathcal{F}_{\mathrm{H}}$ & \multicolumn{1}{c|}{0.995} & \multicolumn{1}{c|}{0.9999} & 0.999 \\
$\mathcal{F}_{\mathrm{Shuttling}}$ & \multicolumn{1}{c|}{1} & \multicolumn{1}{c|}{0.999} & 0.9999 \\ \hline
$t_{\mathrm{U_3}}$ [\SI{}{\micro\s}]& \multicolumn{3}{c|}{0.5} \\
$t_{\mathrm{CZ}}$ [\SI{}{\micro\s}]& \multicolumn{3}{c|}{0.2} \\
$t_{\mathrm{CCZ}}$ [\SI{}{\micro\s}]& \multicolumn{3}{c|}{0.4} \\
$t_{\mathrm{CCCZ}}$ [\SI{}{\micro\s}] & \multicolumn{3}{c|}{0.6} \\ \hline
$v$ [\SI{}{\micro\m\per\micro\s}] & \multicolumn{1}{c|}{0.55} & \multicolumn{1}{c|}{0.2} & 0.3 \\
$t_{\mathrm{act/deact}}$ [\SI{}{\micro\s}] & \multicolumn{1}{c|}{20} & \multicolumn{1}{c|}{50} & 40 \\ \hline
T1 [\SI{}{\micro\s}]& \multicolumn{3}{c|}{100000000} \\
T2 [\SI{}{\micro\s}]& \multicolumn{3}{c|}{1500000} \\ \hline
\end{tabular}
}
\end{subtable}
\end{subtable}
\footnotesize{
\begin{flushleft}
$\Delta \mathrm{CZ}$ and $\Delta T$ represent the difference in the number of CZ gates and circuit execution time, respectively.
$\delta \mathcal{F}$ measures the relative fidelity decrease, taking the negative logarithm of the approximate success probability (less is better).
RT is the runtime of the mapping process in CPU seconds,
$\alpha = \alpha_\mathrm{g}/\alpha_\mathrm{s}$ is the ratio between gate- and shuttling-based mapping.
\end{flushleft}
}
\end{table*}

\subsubsection{Shuttling-based Routing}
\label{sec:shuttl-based-rout}
As discussed in \Cref{sec:increased-search}, considering all possible shuttling operations is unfeasible, and only the operations moving the qubits directly to their destination are taken into consideration.
In general, there are two cases (as discussed in \Cref{ex:search-space}): a direct move $M$ to an unoccupied coordinate, or a move combination $(M_{\mathrm{away}}, M)$ consisting of a first move-away operation, followed by the direct move to the now free coordinate.
These moves are then combined into a chain of movements, listing all shuttling operations required to execute a certain gate, where the chain length is bounded by $2 (m-1)$.
This represents the worst case where all gate qubits have to use a move-away combination to make the gate executable.
These movement chains are created for each gate qubit, always keeping chains of minimal length, based on the intuition that two move operations are unlikely to be faster than a direct single operation, even if they can be shuttled in parallel.
The chains themselves are created by recursively choosing a qubit of the vicinity $V$ of the central gate qubit and considering the possibility of moving all other gate qubits close to this position.
This is done by first choosing qubits that allow for a direct move, resulting in a fast and close-to-optimal selection of interesting move operations.

Each move chain is then evaluated by summing the following cost function over all individual moves $M$ contained in the chain:
\begin{align}
  \label{eq:cost-shuttling-based}
C_{\mathrm{s}}(M) &= C^{f}_{\mathrm{s}}(M) + w_{l} \, C^{l}_{\mathrm{s}}(M) + w^{t}C^{t}_{\mathrm{parallel}}(M) \\
  C^{f}_{\mathrm{s}}(M) &= \sum_{g \in f_{\mathrm{s}}} \Delta d(M), \quad C^{t}_{\mathrm{parallel}}(M) = \sum_{M^{t}} \Delta T(M,M^{t}) \, ,
\end{align}
where the first two terms correspond to a distance reduction in the front and lookahead layer, similar to \Cref{eq:cost-gate-based}.
$C^{t}_{\mathrm{parallel}}$ takes into account if the the moves of the move chain can be executed in parallel to the last $t$ move operations $M^{t}$, i.e.,
\begin{equation*}
  \label{eq:cost-shuttling-parallel}
  \Delta T(M,M^{t}) = \begin{cases}
                                  0  & \text{, parallel loading \& shuttle}\\
                                  t_{\mathrm{act}} + t_{\mathrm{deact}} & \text{, parallel loading}\\
                                  t_{\mathrm{act}} + s(M)/v + t_{\mathrm{deact}} & \text{, else}\\
                                  \end{cases}
\end{equation*}
where $t_{\mathrm{(de)act}}$ is the (de)activation time of the AOD coordinates, $v$~the shuttling velocity, and $s(M)$ the rectangular shuttling distance of movement $M$.
By varying the \emph{time weight} $w^{t}$ one can therefore control the contribution of the parallelism constraint to the overall cost function and, therefore, it plays a similar role to $\lambda^t$ in \Cref{eq:cost-gate-based} controlling the trade-off between choosing the most effective operation versus the one that can be executed best in parallel with previous operations.

For further technical details, we refer to the code which is publicly available at~\cite{LsschmidMqtqmapHybridmapper}.

\section{Numerical Evaluations}
\label{sec:numer-eval}

The advantage of the hybrid mapping approach proposed in this work is its flexibility when it comes to different hardware configurations.
By leveraging SWAP gate insertion and shuttling, the mapper can choose for each gate in the front layer the currently favorable mapping capability with potential improvements regarding circuit runtime and average circuit fidelity.
These benefits have been confirmed in experimental evaluations, with the main results of the evaluations summarized in this section.

\subsection{Experimental Setup}
\label{sec:experimental-setup}

We performed mapping experiments across three different hardware configurations whose corresponding settings are summarized in \Cref{tab:table1_c}.
They correspond to a (1) shuttling-optimized, a \mbox{(2) gate-optimized}, and a (3) mixed configuration that does not have a favorable mapping capability.
The considered hardware size is a $15\times15$ lattice with $d=\SI{3}{\micro\m}$ and $N=200$ atoms in all cases.

As benchmarks, we utilized three representative quantum circuits~\cite{quetschlich2023mqtbench}, namely the \emph{Quantum Fourier Transform} (QFT), \emph{Quantum Phase Estimation} (QPE), and a \emph{graph state preparation circuit} (graph) for $n=200$ qubits.
To account for multi-qubit gates, we additionally considered three benchmark circuits representing classical binary reversible functions synthesized by~\cite{adarsh2022syrec} using $\mathrm{C}_{m}\mathrm{X}$ gates with $m \leq 4$ denoted \emph{bn}, \emph{call}, and \emph{gray} (\Cref{tab:table1_b}).
For all circuits, $\mathrm{C}_{m}\mathrm{X}$ have been decomposed to natively supported $\mathrm{C}_{m}\mathrm{Z}$ gates.

For each hardware and circuit, the mapper is executed in three different modes: (A) gate-based only: $\alpha_\mathrm{s}=0$, (B) shuttling-based only: $\alpha_\mathrm{g}=0$, and (C) hybrid mapping utilizing the full hybrid compilation process proposed in \Cref{sec:hybr-mapp-problem}.
For the hybrid mode, different decision radios $\alpha = \alpha_\mathrm{g}/\alpha_\mathrm{s}$ are tested, keeping only the best.
The remaining mapping parameters are $\lambda^{t}=0$, $w_{l} = 0.1$, $w^{t}=0.1$, and $t=4$. 
A trivial identity mapping is chosen for the initial layout, \mbox{i.e.,~$q_i \leftrightarrow Q_i \leftrightarrow C_i$} for all circuit qubits.

To evaluate the results, both the original and the mapped circuits are scheduled, taking again state-of-the-art hardware parameters of \Cref{tab:table1_c}.
The difference in circuit execution time $\Delta T$ and the number of CZ gates $\Delta \mathrm{CZ}$ are computed, where the latter correspond to the eventually inserted and decomposed SWAP gates.
Additionally, in both cases, a total average mapping fidelity is computed and compared taking the negative logarithm of the approximate success probability $P$ ratio $\delta \mathcal{F} = - \log(P_{\mathrm{mapped}}/P_{\mathrm{original}})$ with a smaller value representing a smaller fidelity decrease caused by the mapping process.
The results with the corresponding mapper runtimes (RT) in CPU seconds are listed in \Cref{tab:table1_a}.

\subsection{Discussion}
\label{sec:discussion}
It is obvious that shuttling-only mapping results in $\Delta \mathrm{CZ}=0$, as no additional gates are added. 
For gate-based mapping, on the other hand, the inserted SWAP gates are decomposed and, therefore, result in a higher $\Delta \mathrm{CZ}$, while the time overhead $\Delta T$ is magnitudes of order smaller compared to the slower shuttling-only mapping.
Nevertheless, for the (1)~\mbox{shuttling-optimized} hardware it is still favorable to utilize the slow shuttling compared to the error-prone CZ gates, due to the long coherence times.
Similarly, for the (2)~\mbox{gate-optimized} hardware with improved CZ fidelity, SWAP gate insertion represents the preferred mapping capability as expected.
The proposed hybrid mapper directly utilizes the hardware parameters to decide between the two capabilities and can, therefore, correctly identify the more suitable strategy, resulting in the best output in all shown cases.

Moreover, the full flexibility of the proposed hybrid approach is demonstrated in the last row of \Cref{tab:table1_a}, corresponding to the (3)~mixed hardware setup, representing reasonable operation fidelities for near-term devices.
Instead of using the same computational capability through the whole mapping process, the hybrid mapper can choose the most suitable way to map each gate within the circuit, resulting in a combination of SWAP gates and shuttling moves.
By leveraging both mapping capabilities the hybrid mapper can reduce the fidelity decrease $\delta \mathcal{F}$ caused by the mapping, for all considered benchmarks except for the graph-state preparation.
Here, the mapper correctly identifies the shuttling-only approach to be still the best choice.
Note that the optimal ratio $\alpha$ between gate- and shuttling-mapping varies for different circuits, indicating a connection between circuit structure and preferred mapping capability.
The proposed hybrid mapper allows, for the first time, to study this correlation, with a systematic case study left for future work.

\section{Conclusions}
\label{sec:summary-outlook}

In this work, we proposed a hybrid compilation methodology, incorporating both gate-based SWAP insertion and \mbox{shuttling-based} qubit array rearrangements, i.e., both mapping capabilities of the Neutral Atom (NA) platform.
We addressed challenges arising from the simultaneous utilization of both mapping capabilities and integrated corresponding solutions into an overall hybrid compilation process, inclusive of direct support for \mbox{multi-qubit} gates.
The versatility of the proposed approach has been demonstrated across diverse hardware configurations, showcasing its adaptability and potential for enhancing fidelity.
By applying the proposed mapping process to quantum circuits featuring up to 200 qubits, we have shown its effectiveness in harnessing the complete spectrum of computational capabilities provided by NAs, encompassing \mbox{high-fidelity} \mbox{long-range} interactions, native \mbox{multi-qubit} gate support, and qubit shuttling.
This opens new possibilities for future studies and compiler development for the NA platform.

\begin{acks}
The authors thank Johannes Zeiher for fruitful discussions and comments regarding Neutral Atoms.

L.S. and R.W. acknowledge funding from the European Research Council (ERC) under the European Union’s Horizon 2020 research and innovation program (grant agreement No. 101001318) and this work was part of the Munich Quantum Valley, which is supported by the Bavarian state government with funds from the Hightech Agenda Bayern Plus,

\end{acks}

\printbibliography

\end{document}